# On the $k_\perp$ dependent gluon density in hadrons and in the photon

J. Blümlein

DESY–Zeuthen, Platanenallee 6, D–15735 Zeuthen, Germany

**Abstract**

The $k_\perp$ dependent gluon distribution for protons, pions and photons is calculated numerically accounting for the resummation of small $x$ effects due to the Lipatov equation. It is represented by a convolution of a gluon density in the collinear limit and a universal function $\mathcal{G}(x, k^2, \mu)$ for which an analytic expression is derived.

# 1 Introduction

In the small $x$ range new dynamical effects are expected to determine the behaviour of structure functions. The evolution of parton densities is effected by terms due to non strong $k_\perp$ ordering and eventually by screening contributions. This behaviour is expected to determine not only the structure functions of the nucleon but also the parton distributions of the pion and the photon. A description of contributions of this type requires to generalize the factorization of the hadronic matrix elements into coefficient functions and parton densities in which the $k_\perp$ dependence is not integrated out [1].

This factorization covers the case of collinear factorization in the limit that the $k^2$ dependence of the coefficient function is neglected. The $k_\perp$ dependent gluon density accounts for the resummation of small $x$ effects. In the present paper we will consider those due to the Lipatov equation only. Since this equation behaves infrared finite no other singularities will emerge than in the case of mass factorization. The collinear singularities are delt with in the same way in the case of $k_\perp$ factorization.

In the present paper the $k_\perp$ dependent gluon distribution is calculated for the case of the scheme [2, 3] for the proton, pion, and the photon. The $k_\perp$ dependent distribution $\Phi(x, k^2, \mu)$ can be represented as the convolution of the gluon density in the collinear limit $g(x, \mu)$ and a function $\mathcal{G}(x, k^2, \mu)$ for which an analytic expression will be derived. A numerical comparison of the $k_\perp$ dependent gluon densities accounting for the solution of the Lipatov equation and in the double logarithmic approximation (DLA) is given for the different hadrons and the photon. It may be interesting to try to unfold these distributions using different observables $(F_2, F_L, \sigma_{2\ jet}, \sigma_{J/\psi})$ and to compare the behaviour of this distribution for different hadrons $(p, \pi)$ and the photon to obtain a conclusive understanding of this quantity. Here we focus on the quantitative behaviour of $\Phi(x, k^2, \mu)$ as a function of $x$ and $k^2$ to investigate first the size of the novel effects in comparison with standard ones.

# 2 $k_\perp$ Factorization and the $k_\perp$ dependent gluon distribution

The factorization relation for an observable $O_i(x, \mu)$ reads

$$O_i(x, \mu) = \int dk^2 \hat{\sigma}_{O_i}(x, k^2, \mu) \otimes \Phi(x, k^2, \mu) \qquad (1)$$

where $\hat{\sigma}_{O_i}(x, k^2, \mu)$ and $\Phi(x, k^2, \mu)$ denote the $k^2$ dependent coefficient function and parton density[1], respectively. Eq. (1) can be rewritten as [2, 3]

$$O_i(x, \mu) = \hat{\sigma}_{O_i}^0(x, \mu) \otimes G(x, \mu) + \int_0^\infty dk^2 \left[\hat{\sigma}_{O_i}(x, k^2, \mu) - \hat{\sigma}_{O_i}^0(x, \mu)\right] \Phi(x, k^2, \mu) \qquad (2)$$

with $\hat{\sigma}_{O_i}^0(x, \mu) = \lim_{k^2 \to 0} \hat{\sigma}_{O_i}(x, k^2, \mu)$. The first addend in (2) describes the conventional contribution due to collinear factorization. The second term contains the new contributions. Note that $\Phi(x, k^2, \mu)$ starts with terms $\propto \overline{\alpha}_s$. It has therefore *not* the interpretation of a probability density and may even become negative.

---

[1] We will consider the gluon density in the present paper only.

As shown in [2] the $k_\perp$ dependent gluon distribution associated to eq. (2) reads in moment space

$$\tilde{\Phi}(j,k^2,\mu) = \gamma_c(j,\overline{\alpha}_s)\frac{1}{k^2}\left(\frac{k^2}{\mu^2}\right)^{\gamma_c(j,\overline{\alpha}_s)}\tilde{g}(j,\mu) \tag{3}$$

where $\mu$ denotes a factorization scale, $\overline{\alpha}_s = N_c\alpha_s(\mu)/\pi$, and $g(x,\mu)$ is the gluon density. Eq. (3) accounts for the small $x$ behaviour due to the Lipatov equation. Here $\gamma_c(j,\overline{\alpha}_s)$ is the solution of the eigenvalue equation of the *homogeneous* Lipatov equation[2]

$$\rho \equiv \frac{j-1}{\overline{\alpha}_s} = \chi(\gamma_c(j,\overline{\alpha}_s)), \quad \chi(\gamma) = 2\psi(1) - \psi(\gamma) - \psi(1-\gamma). \tag{4}$$

In $x$ space the $k_\perp$ dependent distribution is given by the convolution

$$\Phi(x,k^2,\mu) = \mathcal{G}(x,k^2,\mu) \otimes g(x,\mu), \tag{5}$$

correspondingly, with

$$\int_0^{\mu^2} dk^2 \Phi(x,k^2,\mu) = \delta(1-x). \tag{6}$$

The function $\mathcal{G}(x,k^2,\mu)$ is universal and can be calculated numerically by a contour integral in the complex plane over the first factor in eq. (3). Since the solution of eq. (4) is multivalued the Mellin inversion to $x$ space requires to select the branch in which for asymptotic values of $j \in \mathcal{C}$ $\gamma_c$ approaches the perturbative result $\gamma_c(j,\overline{\alpha}_s) \sim \overline{\alpha}_s/(j-1)$ for small values of $\overline{\alpha}_s$.

We solved eq. (4) under this condition numerically using an adaptive Newton algorithm. The solution is characterized by three branch points. Their position in the $\rho$ plane may be determined using the solutions of (7) in eq. (4)

$$\psi'(z) - \frac{\pi^2}{2}\frac{1}{\sin^2(\pi z)} = 0. \tag{7}$$

Obviously $z \equiv \gamma_c^s = 1/2$ is one of the solutions of (7) leading to $\rho_1^s = 4\ln 2$. The two other solutions are

$$\begin{aligned}\gamma_{c(2,3)}^s &= -0.4252 \pm 0.4739\ i \\ \rho_{(2,3)}^s &= -1.4105 \pm 1.9721\ i.\end{aligned} \tag{8}$$

This agrees with a result in ref. [5]. As shown in figure 1 the smooth dependence of $\gamma(\rho)$ for $\text{Re}\rho > 4\ln 2$ changes if $\text{Re}\rho < \text{Re}\rho_1^s$. $\text{Im}\gamma$ becomes discontinuous at $\text{Im}\rho = 0$ and $\text{Re}\gamma$ developes a 'ridge' with $\text{Re}\gamma = 0.5$ at $\text{Im}\rho = 0$. In figure 2 and 3 the real and imaginary part of $\gamma_c$ are illustrated for smaller values of $\text{Re}\rho$. For the real part the 'ridge' persists until $\text{Re}\rho \sim -1.3$ and turns into a flat form with $\text{Re}(\gamma_c) \sim -0.5$ for $|\text{Im}\rho| < 1.5$. At the same time the imaginary part of $\gamma_c$ becomes continuous again. We have also shown the contour of $\text{Re}\gamma_c$ and $\text{Im}\gamma_c$ for $\text{Re}\rho = \text{Re}\rho_{(2,3)}^s$. The branch points are clearly seen as edges in $\text{Re}\gamma_c$ and tips in $\text{Im}\gamma_c$. The curvature of $\text{Re}, \text{Im}(\gamma_c)$ changes its sign upon passing the branch point at $\text{Re}\rho = \text{const.}$ discontinuously. At slightly larger values of $\text{Re}\rho$ this change is *continuous* still (cf. figure 2,3).

---

[2]In several recent approaches [4] to describe $k_\perp$ dependent gluon distributions phenomenological Ansätze were used based on solutions of *inhomogeneous* Lipatov equations. Note that these descriptions are not related to eq. (2) and [2, 3].

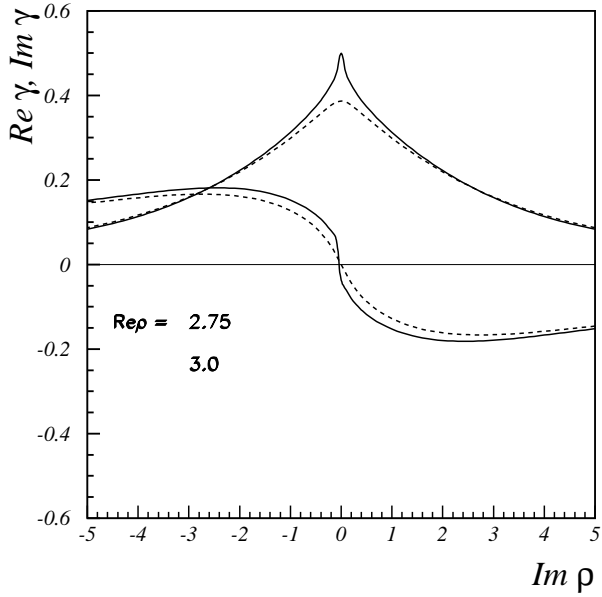

Figure 1: Real and imaginary part of $\gamma_c$ in the vicinity of the branch point at $\rho_1^s = 4\ln 2$.

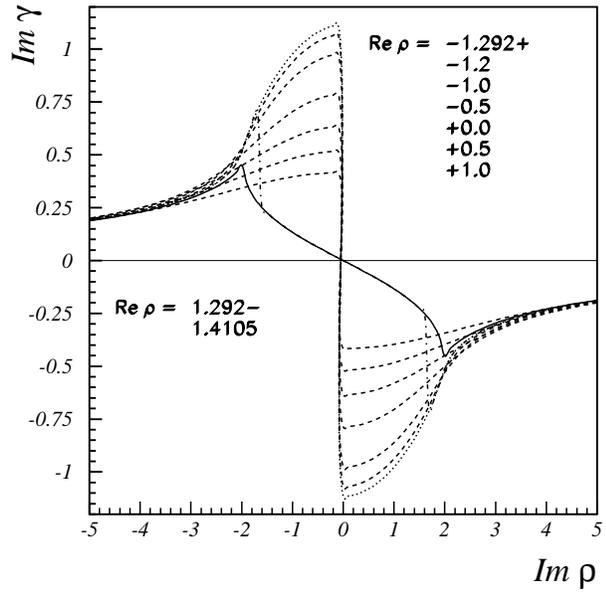

Figure 3: Imaginary part of $\gamma_c$ in the range $\text{Re}\rho \in [\text{Re}\rho_{(2,3)}^s, 1]$.

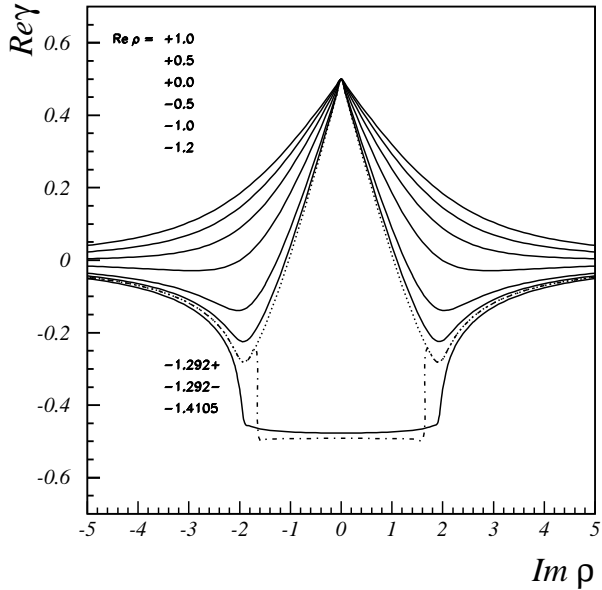

Figure 2: Real part of $\gamma_c$ in the range $\text{Re}\rho \in [\text{Re}\rho_{(2,3)}^s, 1]$.

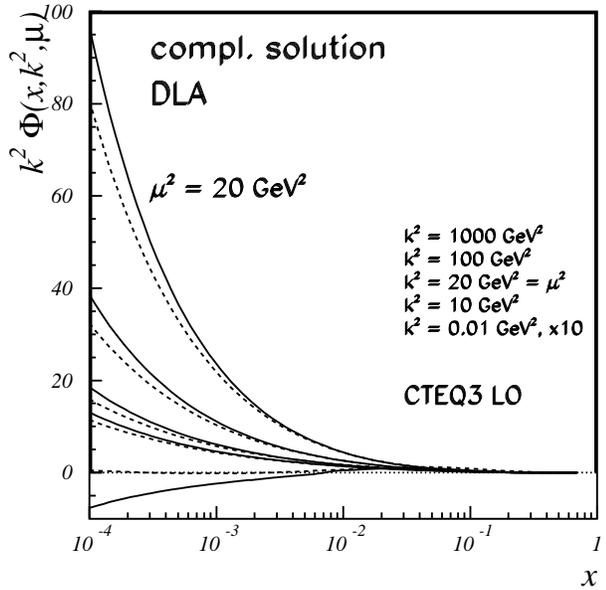

Figure 4: The $k_\perp$ dependent gluon distribution of the proton as a function of $k^2$ and $x$. Full lines: complete solution; dashed lines: solution in DLA. For the input distribution $g(x, \mu)$ the parametrization [8] (LO) was used.

# 3 An analytical solution for $\mathcal{G}(x, k^2, \mu)$

The integration contour for the Mellin transformation of $\tilde{\Phi}(j, k^2, \mu)$ to $\Phi(x, k^2, \mu)$ for $j \in \mathcal{C}$ has to be situated outside the range of the singularities of $\gamma_c$. One may expand $\gamma_c(j, \overline{\alpha}_s)$ into a Laurent series over $\rho$

$$\gamma_c(j, \overline{\alpha}_s) = \sum_{l=1}^{\infty} g_l \rho^{-l} \qquad (9)$$

in this range.

The coefficients $g_l$ are given in [6] up to $l = 20$ in analytical form extending an earlier result [7]. For small values of $|\mathrm{Im}\rho|$ ($|\mathrm{Im}\rho| < 2$) the truncated Laurent series leads to an oscillatory behaviour and eq. (9) does no longer serve to be an appropriate description of $\gamma_c$ (cf. [6]).

Using (9) a corresponding expansion may be performed for

$$k^2 \tilde{\mathcal{G}}(j, k^2, \mu) = \gamma_c(j, \overline{\alpha}_s) \exp[\gamma_c(j, \overline{\alpha}_s) L] \qquad (10)$$

with $L = \ln(k^2/\mu^2)$. For the single terms of the Laurent series in $\rho$ the Mellin transform can be carried out analytically.

Here it is important to expand the exponential in eq. (10) in such a way that the lowest order term in $\overline{\alpha}_s$ of $\gamma_c$ is *kept* in exponential form. One obtains

$$\begin{aligned} k^2 \mathcal{G}(x, k^2, \mu) &= \frac{\overline{\alpha}_s}{x} I_0 \left(2\sqrt{\overline{\alpha}_s \log(1/x) L}\right) + \frac{\overline{\alpha}_s}{x} \sum_{\nu=4}^{\infty} d_\nu(L) \left(\frac{\overline{\alpha}_s \log(1/x)}{L}\right)^{(\nu-1)/2} \\ &\quad \times I_{\nu-1} \left(2\sqrt{\overline{\alpha}_s \log(1/x) L}\right), \; L > 0. \end{aligned} \qquad (11)$$

The coefficients $d_\nu(L)$ are given in ref. [6]. Up to $\nu = 20$ they contain at most terms $\propto L^4$. The first term in eq. (11) denotes the Green's function in DLA.

For $L \to 0$ (11) takes the form

$$k^2 \mathcal{G}(x, k^2, \mu) = \frac{\overline{\alpha}_s}{x} \sum_{l=1}^{\infty} \frac{g_l}{(l-1)!} \left[\overline{\alpha}_s \left(\frac{1}{x}\right)\right]^{l-1}, \qquad (12)$$

and for $L < 0$ (i.e. $k^2 < \mu^2$) one has

$$\begin{aligned} k^2 \mathcal{G}(x, k^2, \mu) &= \frac{\overline{\alpha}_s}{x} J_0 \left(2\sqrt{\overline{\alpha}_s \log(1/x) |L|}\right) + \frac{\overline{\alpha}_s}{x} \sum_{\nu=4}^{\infty} d_\nu(L) \left(\frac{\overline{\alpha}_s \log(1/x)}{|L|}\right)^{(\nu-1)/2} \\ &\quad \times J_{\nu-1} \left(2\sqrt{\overline{\alpha}_s \log(1/x) |L|}\right). \end{aligned} \qquad (13)$$

Thus for $k^2 \to 0$ damped, oscillating modes are obtained which vanish faster than $1/|L|^{-1/4}$.

# 4 Numerical Results

The $k_\perp$ dependent gluon distribution $\Phi(x, k^2, \mu)$ (scaled by $k^2$) is shown in figure 4 as a function of $x$ and $k^2$ for $\mu^2 = 20 \text{ GeV}^2$ for the proton referring to the parametrization of ref. [8] to describe $g(x, \mu)$. The complete solution eq. (5) is larger than the DLA result for $k^2 \gtrsim \mu^2$ at $x \lesssim 10^{-3}$ by 10 to 15% while for $k^2 \to 0$ smaller values are obtained.

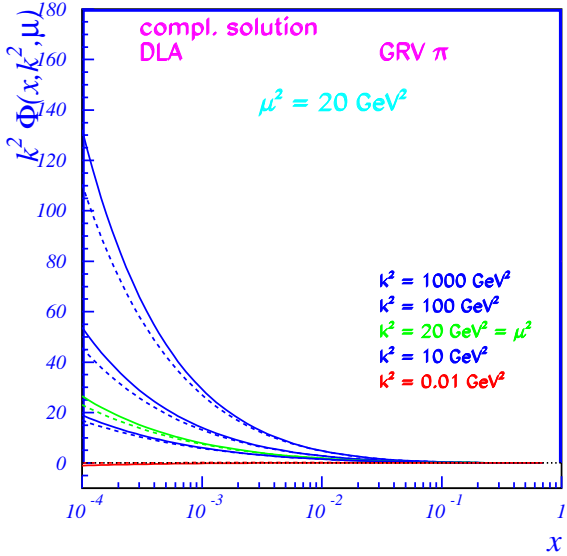 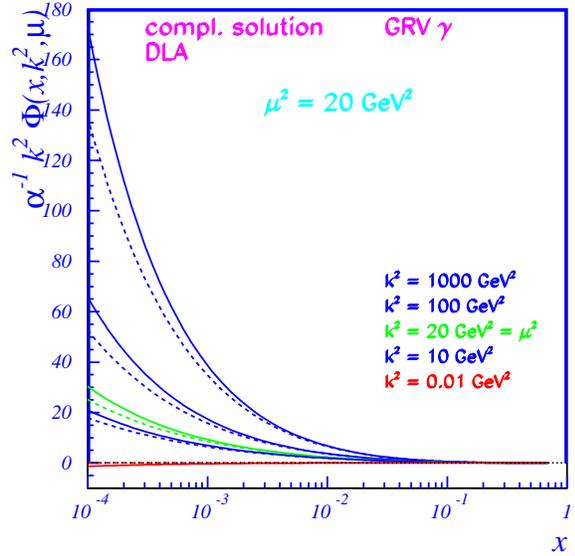

Figure 5: The $k_\perp$ dependent gluon distribution of the pion as a function of $k^2$ and $x$. Full lines: complete solution; dashed lines: solution in DLA. For the input distribution $g(x,\mu)$ the parametrization [9] (LO) was used.

Figure 6: The $k_\perp$ dependent gluon distribution of the photon scaled by $1/\alpha_{QED}$ as a function of $k^2$ and $x$. Full lines: complete solution; dashed lines: solution in DLA. For the input distribution $g(x,\mu)$ the parametrization [10] (LO) was used.

At larger values of $x$ the complete solution approaches the DLA result. For $k^2 \to 0$ $\Phi(x,k^2,\mu)$ vanishes. Since the DLA result is proportional to $J_0(2\sqrt{\overline{\alpha}_s \log(1/x) \log|k^2/\mu^2|})$ for $k^2 \to 0$ a damped oscillatory behaviour is obtained in this approximation. The complete solution, on the other hand, behaves monotonous in the whole kinematical range.

The calculation of $\mathcal{G}(x,k^2,\mu)$ in eq. (5) by numerical Mellin inversion using the numerical solution of eq. (4) is rather time consuming compared to the convolution of the analytical solution (sect. 3) with the input distribution $g(x,\mu)$. We compared both methods and found that the representation given in the previous section leads to a relative error of less than 0.002 using an expansion up to $O(\overline{\alpha}_s^{20})$.

The same calculation has been performed for the $k_\perp$ dependent gluon distribution of the pion and the photon using the distributions [9, 10] (LO) as input parametrizations. The ratio of the complete solution and the solution in DLA is of similar size than in the case of the proton. Comparing the functions $\Phi_\pi(x,k^2,\mu)$ and $\alpha_{QED}^{-1}\Phi_\gamma(x,k^2,\mu)$ the latter one turns out to be larger at virtualities $k^2 \gtrsim \mu^2$.

Since shape and size of the complete solution and in DLA are rather similar very precise measurements are required to establish the non–DLA contributions at small $x$ both in the case of hadrons $(p,\pi)$ and photons.

## 5  Conclusions

We have calculated the $k_\perp$ dependent gluon density for the proton, pion, and photon numerically in leading order using the BFKL equation. A consistent treatment of observables is possible in the scheme [2, 3].

The Green's function $\mathcal{G}(x, k^2, \mu)$ was found both numerically and by a perturbative analytic expression expanding the complete solution up to $O(\overline{\alpha}_s^{20})$. Both representations agree better than 0.002 after convoluting with the respective input distributions $g(x, \mu)$.

The effect of the non–DLA terms in $\Phi(x, k^2, \mu)$ is of $O(10...15\%)$. To reveal these contributions requires very accurate measurements in the small $x$ range in the case of *every* observable being sensitive to the gluon density widely independent of the choice of the target.

# On the $k_\perp$ dependent gluon density in hadrons and in the photon

Johannes Blümlein

*DESY–Zeuthen, Platanenallee 6, D–15735 Zeuthen, Germany*

**Abstract**

The $k_\perp$ dependent gluon distribution for protons, pions, and the photon. is calculated accounting for the resummation of small $x$ effects due to the Lipatov equation. It is represented by a convolution of a gluon density in the collinear limit and a universal function $\mathcal{G}(x, k^2, \mu)$ for which an analytic expression is derived.